\documentclass[fleqn,twoside]{article}
\usepackage[headings]{espcrc2}

\readRCS
$Id: espcrc2.tex,v 1.2 2004/02/24 11:22:11 spepping Exp $
\usepackage{graphicx}
\usepackage[figuresright]{rotating}

\newcommand{\AmS}{{\protect\the\textfont2
  A\kern-.1667em\lower.5ex\hbox{M}\kern-.125emS}}

\title{Chiral Dynamics beyond the Standard Model}

\author{Jan Stern\address{Theory Group, IPN-Orsay, Universite de Paris Sud XI, 
        91406 ORSAY, France }}

\runtitle{Chiral Dynamics beyond the SM}
\runauthor{J. Stern}

\begin{document}

\begin{abstract}
The SM Lagrangian whithout physical scalars is rewritten as the LO of a
Low-Energy Effective Theory invariant under a higher non linear symmetry
$S_{nat}\supset SU(2)_W \times U(1)_Y$. Soft breaking of $S_{nat}$ defines
a hierarchy of non standard effects dominated by universal couplings of right 
handed quarks to W. The interface of corresponding EW tests with non
perturbative QCD aspects is briefly discussed.
\vspace{1pc}
\end{abstract}

\maketitle

\section{ELECTROWEAK LOW-ENERGY EFFECTIVE THEORY}

A systematic search of  Physics beyond the Standard Model is usually
formulated in the framework of a Low-Energy Effective theory (LEET). One expects
that at very high energies $E > \Lambda$ there exist new (gauge) particles
and that their interaction is governed by new (local) symmetries not contained
in the SM gauge group $SU(2)_W \times U(1)_Y$. In order to specify the LEET
which would describe the Physics below the scale $\Lambda$ in a systematic
low-energy expansion, one has to identify the characteristic property 
which makes from the SM the  unique and precise effective description of
low-energy phenomena. The most popular approach is provided by the

\subsection{Decoupling Scenario}

One assumes that below the scale $\Lambda$, it is possible to ignore
(integrate out) the heavy states and forget the corresponding (gauge)
symmetries beyond $SU(2)_W\times U(1)_Y$, i.e. to remain with the     
degrees of freedom
and symmetries of the SM. The latter is then identified as the most general
$SU(2)_W\times U(1)_Y$ invariant Lagrangian that can be constructed out of the
SM fields and is {\bf renormalizable order by order in powers of coupling 
constants}. The effective Lagrangian then reads
\begin{equation}
\mathcal{L}^{eff}=\mathcal{L}_{SM} + \sum_{D>4}\frac{1}{\Lambda^{D-4}}
\mathcal{O}_D
\end{equation}
where D denotes the mass (ultraviolet) dimension of the operator
$\mathcal{O}_D$. Using the UV dimension D as the sole indication of the
degree of suppression of a non standard operator is  related to
the emphasize put on  renormalizability. However, it is not very
practical:  At the NLO order, there are 80 independent gauge invariant D=6
operators \cite{BW} and a finer classification of new operators might be
needed .

\subsection{A not quite decoupling alternative}

Even if below the scale $\Lambda$ the heavy particles decouple, their
interaction may be such that some high energy symmetries $S_{nat}$
beyond $SU(2)_W\times U(1)_Y$ survive at low energies and constrain the LEET.
Since below the scale $\Lambda$ the set of effective fields - typically
the SM fields - is too small to span a linear representation of $S_{nat}$,
the extra symmetry relevant at low energy
\begin{equation}
\frac{S_{nat}}{SU(2)_W\times U(1)_Y}\equiv \mathcal{C}_{sp}
\end{equation}
should be realized nonlinearly. This means that the symmetry (2) is not
manifest in the low-energy spectrum but it constrains interaction vertices.
The effective Lagrangian should exhibit the symmetry $S_{nat}$ and should be
organized in a systematic low-energy expansion in powers of momenta
\begin{equation}
\mathcal{L}^{eff} = \sum_{d\ge 2} \mathcal{L}_d
\end{equation}
where $\mathcal{L}_d =\mathcal{O}(p^d)$ in the low-energy limit.( In general,
the chiral dimension $d\neq D$ . A concise review of infrared power counting
can be found in \cite{Wudka} , \cite{HS2}.) The resulting
non decoupling LEET is {\bf renormalized and unitarized order by order
in the momentum expansion (3)}, following and extending the example of
Chiral Perturbation Theory (ChPT), \cite{Wein} \cite{GL}. Under these circumstances
there is no particular reason to start the LE expansion around a
renormalizable limit: In its minimal version, the LEET may contain all
the observed particles of the SM without the physical Higgs scalar, unless
such a light scalar is found experimentally. On the other hand, the presence
of three GB fields $\Sigma \in SU(2)$ is crucial to generate the masses of W
and Z by the standard Higgs mechanism. The standard gauge boson mass term
coincides with the Goldstone boson kinetic term
\begin{equation}
\mathcal{L}_{mass} =\frac{1}{4} F^2_W Tr(D_{\mu}\Sigma^{\dagger}D^{\mu}\Sigma) .
\end{equation}
(This becomes manifest in the physical gauge $\Sigma=1$). An important
difference between the decoupling and non decoupling LEET concerns the
scale $\Lambda$ above which the effective description breaks down.
 Whereas in the decoupling case $\Lambda$ is essentially independent 
of the low-energy dynamics itself ( the latter is renormalizable) and it
cannot be estimated apriori, the
consistency of the loop expansion in the decoupling case requires
\begin{equation}
\Lambda \approx \Lambda_W = 4\pi F_W \approx 3 TeV.
\end{equation}

\subsection{What is $S_{nat}$?}

The experimental fact that at the leading order of the low-energy expansion
(3) one meets the bare SM interaction vertices should now follow from the
higher symmetry
\begin{equation}
S_{nat}\supset S_{EW}\equiv SU(2)_W\times U(1)_Y
\end{equation}
and not from the requirement of renormalizability as in eq (1) above. Indeed,
as shown in ref \cite{HS1} , given the set of SM fields observed at low energy
and their standard transformation properties under $S_{EW}$, one can contruct
several invariants of $S_{EW}$ carying the  leading infrared dimension $d=2$
that are not present in the SM Lagrangian.
Such operators are not observed and it is a primary role
of the symmetry $S_{nat}$ to suppress them. This presumes that $S_{nat}$ is
a non linearly realized (hidden) symmetry of the SM itself and it should
be possible to infer it from the known SM interaction vertices. This turns out
to be the case \cite{HS1} , \cite{HS2}, provided one sticks to the part
$\mathcal{L}^{\prime}_{SM}$ of the SM Lagrangian that does not contain the
physical Higgs scalar nor the Yukawa couplings to fermions. Proceeding by
trial and error one can show that the condition
$\mathcal{L}_2 = \mathcal{L}^{\prime}_{SM}$
admits a unique {\bf minimal solution}
\begin{equation}
S_{nat} = [SU(2)\times SU(2)]^2 \times U(1)_{B-L}.
\end{equation}
This provides a guide of constructing the LEET below the scale $\Lambda_W$
and it might indicate which new heavy particles could be expected well above
this scale.

\subsection{Spurions}

It is conceivable that at ultrahigh energies the symmetry $S_{nat}$ is
linearly realised via 4 $SU(2)$ and one $U(1)$ gauge fields with 9 of
them acquiring a mass $ >\Lambda_W$. As the energy decreases below
$\Lambda_W$, this linearly realized symmetry is reduced ending up with
$S_{EW}=SU(2)_W\times U(1)_Y$ spaned by the electroweak gauge bosons.
This reduction may be viewed as a pairwise identification of two independent
$SU(2)$ factors in (7)     {\bf up to a gauge transformation $\Omega(x)$}.
Denoting the corresponding $SU(2)_{I}\times SU(2)_{II}$ connections       
by $A^{I}_{\mu}$
and $A^{II}_{\mu}$ respectively, the identification amounts to the constraint
\begin{equation}
A^{I}_{\mu}=\Omega A^{II}_{\mu}{\Omega}^{-1}+ i\Omega \partial_{\mu}{\Omega}^{-1}
\end{equation}
This constraint is covariant under the original symmetry         
$SU(2)_{I}\times SU(2)_{II}$
provided the gauge transformation $\Omega(x)$ is promoted to a field
transforming with respect to the $SU(2)_{I}\times SU(2)_{II}$ group as
the bifundamental representation
\begin{equation}
\Omega(x)\to V_{I}(x) \Omega(x){V_{II}(x)}^{-1}.
\end{equation}
The field $\Omega(x)$ ( more precisely its multiple) is a remnant of the 
reduction procedure and we call it {\bf spurion}: It does not propagate since
its covariant derivatives vanishes $D_{\mu}\Omega(x) = 0$   by virtue of
eq.(8).   There exists
a gauge in which each spurion reduces to a constant multiple of the unite
matrix. After the reduction of the original symmetry $S_{nat}$ to $S_{EW}$,
one remains with the 4 SM gauge fields and with 3 $SU(2)$ valued scalar 
spurions populating the coset space 
\begin{equation}
\mathcal{C}_{sp} = \frac{S_{nat}}{S_{EW}} = [SU(2)]^3
\end{equation}
and transforming as bifundamental representations of the symmetry group  
$S_{nat}$ , as illustrated by the above example,c.f.(9). We refer the reader to
\cite{HS2}, where more details of this construction, quantum number matching
and unicity are given and discussed.

\subsection{Summary}

Before the symmetry $S_{nat}$ is reduced via spurions, the content of the
LEET consists of two disconnected sectors: First {\bf the elementary sector}
governed by the gauge group  $[SU(2)_L\times SU(2)_R\times U(1)_{B-L}]_{el}$
that acts on a set of (elementary) chiral fermion doublets transforming as
[1/2,0;B-L] and [0,1/2;B-L] respectively, (similarly as in LR symmetric models    
\cite{LR})  . Second,
{\bf the composite sector} containing three GBs $\Sigma\in SU(2)$ arising
from the spontaneous breakdown of the ``composite'' symmetry
$[SU(2)_L\times SU(2)_R]_c$ similarly to what happens in QCD. The covariant
constraints  reducing  $S_{nat}\to S_{EW}$ first identify (up to a gauge)
the elementary and the composite $SU(2)_L$. This gives rise to a single
$SU(2)_W$ , the one of the SM , and to a spurion $\mathcal{X}$
transforming as [1/2,1/2]
with respect to the composite $\times$ elementary $SU(2)_L$. Analogously, in the
right handed sector, one identifies the two $SU(2)_R$ groups (spurion
$\mathcal{Y}$)
and finally, the resulting right isospin is identified with $U(1)_{B-L}$
(embeded into $SU(2)$) (spurion $\mathcal{Z}$). What remains is $U(1)_Y$ where
the hypercharge  $Y/2 = I^3_R + (B-L)/2$ as required by the SM. Notice that
the spurion $\mathcal{Y}$ admits a gauge invariant decomposition
$\mathcal{Y} =\mathcal{Y}_u + \mathcal{Y}_d $ making appear projectors
of the up and the down components of right handed doublets.

Hence,  the structure of the SM vertices leads to   three
$SU(2)$ valued spurions which in the physical gauge reduce to three
parameters $\mathcal{X} \to \xi$ ,  $ \mathcal{Y} \to \eta$  and
$ \mathcal{Z} \to \zeta$.
They are external to the SM and they parametrize the small explicit breaking of
the symmetry $S_{nat}$. The fact that the spurion $\mathcal{Z}$ necessarily
carries a non zero value of $B-L$ means that the lepton number violation
is an unavoidable byproduct of the above construction \cite{HS2}: Its strength
is tuned by the parameter $\zeta \ll \xi$ , $ \eta \ll 1$ and it cannot be
anticipated within the LEET alone.

\section{LEADING EFFECTS BEYOND THE STANDARD MODEL}

The whole construction of $\mathcal{L}^{eff}$ can now be deviced in three steps:
\begin{itemize}
\item  One collects all invariants under $S_{nat}$ made up from the   
       corresponding 13 gauge fields, GB field $\Sigma$, chiral fermions and
       the three spurions .
\item  One orders them according to the increasing chiral dimension $d$ and to  
       the number of spurion insertions.
\item  One imposes the invariant constraints between gauge fields and spurions
       eliminating the redundant degrees of freedom (c.f. (8)).
\end{itemize}
By construction, the leading term with $d=2$ and no spurion insertion
coincides with the higgsless vertices of the SM with massive W and Z and
massless fermions. The genuine effects beyond SM are identified with
terms explicitly containing spurions. The latter naturally appear in
a hierarchical order given by the power of the (small) spurionic parameters
$\xi$ , $ \eta $   (we can consistently disregard the tiny spurion $\zeta$ and
LNV processes.) Despite the present unability of the LEET formalism to
anticipate the actual size of individual non standard effects, we might      
well be able to order these effects i.e.,{\bf to predict their relative    
importance}.

\subsection{Fermion Masses and power counting} 

Spurions are needed to construct a $S_{nat}$ invariant fermion mass term:
At leading order such terms are proportional to the operators
\begin{equation}
\mathcal{L}_{fm}= \bar \Psi_L\mathcal{X}^{\dagger} \Sigma \mathcal{Y}_a \Psi_R
\end{equation}
where $a=u $ , $ d$. The chiral dimension of these operators is $d=1$, (1/2 for
each fermion field) and they are furthermore suppressed by the spurionic
factor $\xi \eta$. The mass term of the heaviest fermion (the top quark)
with the chirally
protected mass should count at low energies as $\mathcal{O}(p^2)$. This
suggests the following counting rule for the spurion parameters
\begin{equation}
\xi \eta \sim m_{top}/\Lambda_W \sim \mathcal{O}(p)
\end{equation}
attributing to both $\xi$ and $\eta$ the chiral dimension $d = 1/2$. This
leaves space for the existence of much lighter fermions with
the mass term containing additional powers of $\xi$ and $\eta$.

On the other hand, the understanding of {\bf the smallness of neutrino masses}
within the present LEET framework represents an alternative to the see-saw
mechanism. The Majorana mass term necessarily involves the $\Delta L=2$
spurion $\mathcal{Z}$ which brings in a new (tiny) scale related to LNV.
Yet, one has to find the symmetry that suppresses the neutrino Dirac mass
of the type (11). In \cite{HS1} it has been proposed to associate this latter
suppression with the discrete reflection symmetry
\begin{equation}
\nu_R^i \to -\nu_R^i .
\end{equation}
This is possible since at the leading order the right handed neutrino does
not carry any gauge charge and the symmetry (13) does not prevent the right
handed neutrino
to develope its own (small) Majorana mass. This mechanism of suppression
of neutrino masses has yet another consequence. It forbids {\bf the charged
right-handed leptonic currents} despite the fact that $\nu_R$ remains light.
This corollary will take its importance later.

\subsection{The full NLO}

Sofar, all terms had the total chiral dimension $d=2$ , including (in the case
of the mass term) the spurion factor according to the counting rule (12).
This is characteristic of the LO. There is no $d=3$ term not containing
spurions. On the other hand there are two and only two operators quadratic
in spurions with the total chiral dimension $d=3$. They represent non
standard (i.e. containing spurions) $S_{nat}$ invariant couplings of         
fermions to gauge bosons. In the case of left handed fermions the unique
such operator reads
\begin{equation}
\mathcal{O}_L = \bar \Psi_L \mathcal{X}^{\dagger}\Sigma
\gamma^{\mu} D_{\mu}\Sigma^{\dagger}\mathcal{X}\Psi_L ,
\end{equation}
whereas in the right handed sector the corresponding operator has four
components ($a,b = u,d$) that are separately invariant under $S_{nat}$
\begin{equation}
\mathcal{O}_R^{a,b} = \bar\Psi_R \mathcal{Y}^{\dagger}_{a} \Sigma^{\dagger}
\gamma^{\mu}D_{\mu}\Sigma \mathcal{Y}_{b}\Psi_R .
\end{equation}
Both operators are suppressed by a quadratic spurion factor: In view of the
rule (12), they count as $\mathcal{O}(p^2 \xi^2) \sim\mathcal{O}(p^3)$
and $\mathcal{O}(p^2 \eta^2)\sim\mathcal{O}(p^3)$, repectively. It is
remarkable that the operators (14) and (15) cannot be generated by loops.
This follows from the Weinbergs power counting formula
\begin{equation}
d = 2 + 2 L + \sum_{v} (d_v - 2)
\end{equation}
originally established in the case of ChPT \cite{Wein} and subsequently extended to 
the case of LEET containing massive vector particles, chiral fermions 
\cite{Wudka} as well as spurions \cite{HS1}. Eq (16) represents the chiral
dimension of a connected Feynman diagram containing L loops and the vertices v
of a chiral dimension $d_v \ge 2$. In the low-energy expansion the tree
diagrams with a single insertion of a vertex (14) or (15) are more important
than any loop contribution or than higher order purely bosonic trees.
Accordingly, the NLO operators (14) and (15) are universal in the flavour
space.  Loops,
oblique corrections , flavour dependence (and related FCNC), only come at
the NNLO.
The two operators (14) and (15) thus describe the potentially most important
effects beyond
the SM,  {\bf predicted} in the present framework of a non decoupling LEET.
Unfortunately, it is not straightforward to translate this qualitative
prediction into a more quantitative statement (see \cite{BOPS},
\cite{MI}). The operators (14) and (15) both contain couplings of
fermions to $W$ and to $Z$. The latter involves too many apriori unknown
LECs, especially in the right handed sector (15). For this reason, we first
consider

\subsection{Couplings of quarks to W at NLO}

In the physical gauge and in a flavour basis in which both the mass matrix
of u and d quarks are diagonal, the $LO + NLO$ of a universal charged
current coupled to  W as defined by eqs (14) and (15) reads
\begin{equation}
\mathcal{L}_{CC}=g \big [l_{\mu} + \frac{1}{2} \bar U \big (\mathcal{V}_{eff} \gamma_{\mu}
+ \mathcal{A}_{eff} \gamma_{\mu} \gamma_5 \big ) D\big ] W^{\mu} + hc,
\end{equation}
where $l_{\mu}$ stands for the standard leptonic charged $V-A$ current not
affected at NLO. The matrix notation is used in the flavour (family) space:
$U^{T} = (u ,c , t)$ , $D^{T} = (d ,s ,b)$, whereas
$\mathcal{V}_{eff}$ and $\mathcal{A}_{eff}$ are complex $3\times 3$      
effective EW coupling matrices. At NLO they take the form
\begin{equation}
\mathcal{V}_{eff}^{ij}=(1+ \delta) V_L^{ij} + \epsilon V_R^{ij}\\
\end{equation}
\begin{equation}
\mathcal{A}_{eff}^{ij}= - (1+\delta)V_L^{ij} + \epsilon V_R^{ij},
\end{equation}
where $V_L=L_u L_d^{\dagger}$ and $V_R=R_u R_d^{\dagger}$ with $L_u$,$R_u$ and
$L_d$ ,$R_d$ denoting the pairs of unitary matrices that diagonalize the
masses  of the u-type and d-type quarks respectively.
 The parameters
$\delta = \mathcal{O}(\xi^2)$ and $\epsilon = \mathcal{O}(\eta^2)$ originate
from the spurions. Their magnitude is estimated to be of at most a fraction
of per cent\cite{HS2}. Hence at NLO, the LEET predicts two major non standard 
effects concerning the coupling of quarks to W:

i) The existence of {\bf direct couplings of right handed quarks to W}
( keeping in mind that the discrete symmetry (15) forbids similar couplings
for leptons.)

ii) The chiral generalization of CKM unitarity and mixing.

At the LO one has $\delta = \epsilon = 0$ and one recovers the CKM unitarity
of the SM: $\mathcal{V}_{eff} = - \mathcal{A}_{eff} = V_{CKM}$. At the NLO,
the two distinct matrices $V_L$ and $V_R$ are both unitary (for the same
reason as in the SM) but the effective vector and axial-vector matrices
$\mathcal{V}$ and $\mathcal{A}$ which are accessible in semi-leptonic
transistions are not unitary anymore.

\section{INTERFACE OF THE EW AND QCD EFFECTIVE COUPLINGS }

A  measurement of effective EW couplings $\mathcal{V}_{eff}^{ij}$ and
$\mathcal{A}_{eff}^{ij}$ requires an independent
knowledge of the involved non perturbative QCD parameters like the decay
constants $F_{\pi}$,$F_K$, $F_D$, $F_B$ or the transition form factors such as
$f_{+}^{K_0\pi^{-}}(0) \ldots $. The longstanding problem of an accurate extraction of
the $CKM$ matrix element $V^{us}$ and the related test of the ``$CKM$ unitarity''
 illustrates this point and it becomes even more accute in the presence
of non standard EW couplings, such as RHCs. The unfortunate circumstance is that the most
accurate experimental information on QCD quantities mentioned above ,in turn
come from semi leptonic transitions of the type $P \to l\nu$ and
$P' \to P l\nu$ where $P={\pi, K, D, B}$
and, consequently, the result of their measurement depends on (apriori
uknown) EW couplings (18) , (19) . Finding an exit from this {\bf circular
trap}
is a major task of phenomenological Flavor Physics.

\subsection{Non Standard EW Parameters in the Light Quark Sector}

Let us first concentrate on light quarks $u$, $d$, and $s$. For them the
SM loop effects simulating RHCs are strongly suppressed by at least two
powers of mass. 

Since $V_L^{ub}$ is negligible and
\begin{equation}
\mathcal{V}_{eff}^{ud} = 0.97377(26)\equiv cos\hat{\theta}  
\end{equation}
is very precisely known \cite{MS} from nuclear
$0^{+} \to 0^{+}$ transitions , the light quark effective couplings
$\mathcal{V}_{eff}^{ua}$ , $\mathcal{A}_{eff}^{ua}$ , $a=d ,s$ can be expressed
in terms of three non standard effective EW parameters:  the spurion       
parameter $\delta$
defined in (20),(21) and two RHCs parameters $\epsilon_{NS}$ and
$\epsilon_{S}$ defined as
\begin{equation}
\epsilon V_R^{ud} = \epsilon_{NS } cos\theta   ,  \epsilon V_R^{us} =
\epsilon_{S}  sin\theta ,
\end{equation}
where,upon  neglecting $V_L^{ub}$, we have denoted
\begin{equation}
V_L^{ud} = cos \theta = cos \hat{\theta} ( 1 - \delta - \epsilon_{NS}) ,
V_L^{us} = sin \theta
\end{equation}
In the limit of SM all spurion NS parameters vanish ,
$\delta = \epsilon_{NS} = \epsilon_{S} = 0$ and all light quark
EW effective couplings are fixed by the experimental value
of $\mathcal{V}^{ud}_{eff}$.   

\subsection{$F_{\pi}$ ,$ F_K$ ,$ f_{+}(0)$  and $\mathcal{V}^{us}_{eff}$}

Here $F_{\pi}$ , $F_K$ and $f^{K^0}_{+} (0)$ stand for radiatively corrected
genuine QCD quantities defined as residues of GB poles in two-point and
three-point functions of axial and vector currents. These are the quantities
that are subject to ChPT and/or lattice studies. It is  further understood that
all
isospin breaking effects due to $m_d - m_u$ are included. From the        
experimentally well known branching ratios \cite{KLOE}
$K (l2(\gamma))/\pi (l2(\gamma))$ one can infer
$\mathcal{A}^{us} F_K/\mathcal{A}^{ud} F_{\pi}$  and from the rate of
$K^0_{l3}$ we can extract the value of $|f^{K^0}_{+}(0) \mathcal{V}^{us}|$.
Assuming the Standard Model couplings of quarks to W this is sufficient to
extract very accurate values that the corresponding QCD quantities would take
in a world with vanishing spurion parameters       
$\delta , \epsilon_{NS},  \epsilon_{S}$.
The latter are denoted by a hat. They read
\begin{equation}
\hat{F}_{\pi} = (92.4 \pm 0.2) MeV ,
\hat{F}_K/\hat{F}_{\pi} = 1.182 \pm 0.007,
\end{equation}
\begin{equation}
\hat{f}^{K^0}_{+} (0) = 0.951 \pm 0.005
\end{equation}
Here the errors merely reflect the experimental uncertainties in the
measured branching ratios. In the presence of NS couplings of quarks to W
the values of genuine QCD quantities extracted from semileptonic BRs are
modified. Neglecting higher powers of spurion parameters $\delta$ and
$\epsilon$ one gets using Eqs (21) and (22)
\hspace{-2cm}

\begin{equation}
F^2_{\pi} = \hat{F}^2_{\pi} (1 + 4 \epsilon_{NS})
\end{equation}
\begin{equation} 
\left(\frac{F_K}{F_{\pi}}\right)^2 = \left (\frac{\hat{F}_K}{\hat{F}_{\pi}}
\right)^2 \frac{1+2(\epsilon_{S}-\epsilon_{NS})}{1+\frac{2}{sin^2\hat{\theta}}
(\delta+\epsilon_{NS})}
\end{equation}
\begin{equation}
\left[f^{K^0\pi^-}_+(0) \right]^2 \! \! = 
\left[ \hat{f}^{K^0\pi^-}_+(0)\right]^2 \! \!
\frac{1-2(\epsilon_{S}-\epsilon_{NS})}{1+\frac{2}{sin^2\hat{\theta}}
(\delta+\epsilon_{NS})}
\end{equation}

The 3 NLO EW parameters $\delta$ , $\epsilon_{NS}$ and $\epsilon_S$ can be
constrained using independent informations on $QCD$ quantities $F_{\pi}$,
$F_K$ or $f_+^{K^0\pi}$ . Such information can originate from lattice simulations or from
ChPT based measurements. As a matter of          
example let us mention the possible determination of $F_{\pi}$  from the
$\pi_0 \to 2 \gamma$ partial width or from precision ${\pi\pi}$ scattering
experiments, which are independent from the standard determination
based on the $\pi l2$ decay rate. Despite a present lack of accuracy, this
way may eventually provide a measurement of $\epsilon_{NS}$ through Eqs
(23) and (25).

The non standard EW parameters allow to infer the NLO deviation from the unitarity of
the effective mixing matrix $\mathcal{V}_{eff}$, which is relevant in the
description of $K_{l3}$ decays. One finds
\begin{equation}
\left[ \mathcal{V}^{ud}_{eff}\right]^2 + \left[\mathcal{V}^{us}_{eff}\right]^2
= 1 + 2 (\delta + \epsilon_{NS} cos^2 \hat{\theta} +             
\epsilon_{S} sin^2 \hat{\theta} )
\end{equation}
Due to the presence of RHCs ,  flavour mixing effects in $F_K/F_{\pi}$
and in $f_{+}^{K_0\pi}$ are no more related as in the SM.

\subsection{Enhancement of  $\epsilon_S$ ?}

As already pointed out the genuine spurion parameters $\delta$ and
$\epsilon$ can hardly exceed  0.01. On the other hand , the unitarity of
the right-handed mixing matrix $V_R$  implies
\begin{equation}
|\epsilon_{NS}|^2 cos^2 \theta + |\epsilon_{S}|^2 sin^2 \theta \le \epsilon ^2.
\end{equation}
Since we live  close to the left-handed world, one has
$sin\theta \sim 0.22$, reflecting the well known hierarchy of left handed
flavour mixing. This in turn implies that  $|\epsilon_{NS}|<\epsilon$ remains tiny.
On the other hand , $|\epsilon_{S}| \le 4.5 \epsilon$ and it can indeed be enhanced to
a few percent level , {\bf provided the hierarchy in right-handed flavour mixing is
inverted} ,i.e. $| V_R^{ud}| \ll| V_R^{us}|$. In \cite{BOPS} it has been shown
that a stringent test involving the EW coupling $\epsilon_S - \epsilon_{NS}$
can be deviced  in $K^L_{\mu 3}$ decay, (see \cite{MI}). On the other hand,
it seems rather difficult to find another clean manifestation of RHCs  driven by
$\epsilon_S$ and the remaining two NLO EW constants $\delta$ and
$\epsilon_{NS}$ are presumably too small
to be reliably detected. These  remarks still hold if the preceeding discussion
is extended to
processes involving the short distance rather than chiral QCD dynamics.

\end{document}